\documentstyle[a4,11pt,epsf,twoside]{article}
\begin{document}
\title{Full Simulation Studies of the Silicon Tracker for 
the Linear Collider Detector
       \thanks{ Presented at the International Workshop on Linear Colliders, 
August 26 - 30, 2002 in  Jeju, Korea. }
}

\author{M. Iwasaki
	\thanks{e-mail address: masako@phys.s.u-tokyo.ac.jp}
\\
\\
       {\it Dept. of Physics, University of Tokyo,}\\
       {\it 7-3-1 Hongo, Bunkyo-ku, Tokyo 113-0011 JAPAN}
}
\date{}
\maketitle
\begin{abstract}
A central tracker based on  silicon microstrip sensors
has been envisaged for the $e^+e^-$ linear collider experiments.
A full simulation program based on GEANT4 has been developed to study
performance of the tracker.
We report some preliminary results obtained using this program.
\end{abstract}

The use of silicon microstrip
detectors as precise tracking devices in  colliding-beam experiments
has been well established.
The silicon microstrip detector is used as 
a vertex detector mounted immediately outside the beam pipe 
to reconstruct the decay vertices of short-lived particles.
The robustness of the silicon vertex detector against
beam backgrounds has also been demonstrated.
At the SLAC Linear Collider (SLC) experiment,
excessive occupancy 
due to the accelerator originated backgrounds
in the central tracking drift chamber limited its operation.
Because the anticipated accelerator-induced background 
of the future $e^+e^-$ linear collider (LC) is severer than that of SLC,
the silicon microstrip detector has emerged as a promising 
candidate for a central tracking device.

In this report, we summarize preliminary results obtained based on a full simulation of
the silicon microstrip central tracker.
For the detector configuration, we use the parameters considered for
the NLC SD silicon tracker\cite{LCD}, consisting of five concentric
double-sided silicon microstrip layers located at radii between 20cm  and 125cm,
immersed in  the magnetic field of 5 Tesla.
The thickness of each layer is assumed to be 0.5\% of one radiation length.
GEANT4\cite{GEANT4} is used to simulate detector hits and generated 
hit positions are smeared by a resolution of 7$\mu$m.
In order to reconstruct the track, the hits belonging to a particle are grouped and the track fitting program
using Kalman filter is applied to the hits.
Throughout the study presented here, we assume the center-of-mass energy of 500GeV.

\begin{figure}
\epsfysize6cm
\center{
\epsfbox{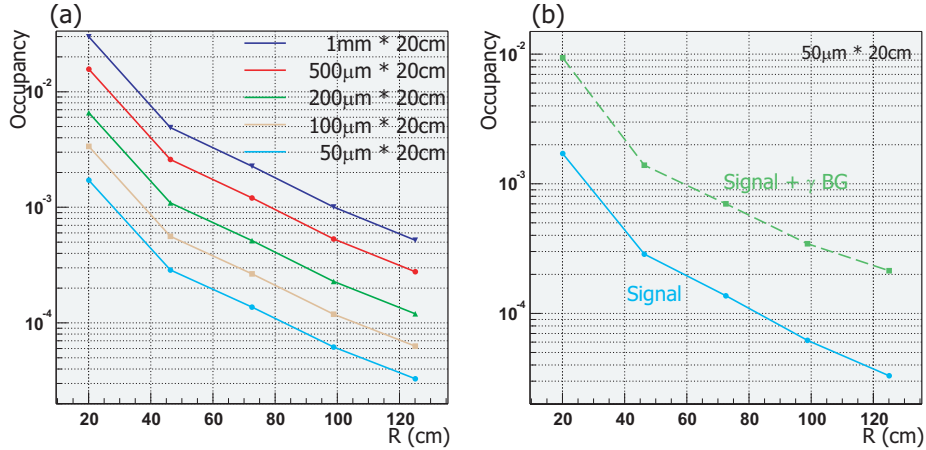}
}
\vskip -0.5cm
\caption{Occupancy distributions as a function 
of the radius, (a) for signal only with different strip-pitch sizes, 
and (b) for signal only (solid line) and  signal plus
photon background at the IR (dashed line).}
\label{fig:1}
\end{figure}
Figure~\ref{fig:1}(a) shows the occupancy of the silicon layer 
as a function of its radial position.
We use the high multiplicity events of
$e^+e^- \rightarrow t\bar{t} \rightarrow$ 6 jets.
The strip length is assumed to be 20 cm while the strip pitch varies
from 50$\mu$m to 1mm.
We find that 
the occupancy of the innermost layer
is 10$^{-3}-10^{-2}$, depending on the pitch size,
for signal-only events.

The simulation of the interaction region (IR) 
indicates that the tracking system is expected to have photon 
backgrounds from unconverted beamstrahlung radiations. 
To estimate the effects caused by the background
we generate beamstrahlung photon backgrounds using the IR simulation,
feed them to the GEANT4 
detector simulation, and superimpose the resulting hits to the signal hits.
For  the NLC accelerator parameter of 190 bunches/train, we find 
$\sim 34000$ photons with the average energy 
of $~1~{\rm MeV}$ are produced in an event. 
Of those produced photons
about 2000 photons with the average energy of 0.7~GeV generate  hits in the tracker.
We find  that one photon generate 1.04 hits on average.
The resulting occupancy is shown as a dashed line 
in Fig.\ref{fig:1}(b). 
The occupancy is an order of magnitude higher than that of 
signal-only events,  
but  still remains  low enough ($\sim 1\%$) even  at the innermost layer.

In order to study the two-track separation capability of the silicon detector,
we generate events with high track density (jet events), 
$e^+e^-\rightarrow q\bar{q}$ ($q$ = $uds$), 
and investigate the minimum distance, within each silicon layer,
between two hits belonging to
different tracks. 
We find the minimum distance of
100$\mu$m, 500$\mu$m,  and 1mm 
in the first,  third and fifth layer, respectively.
\begin{figure}
\epsfysize6cm
\center{
\epsfbox{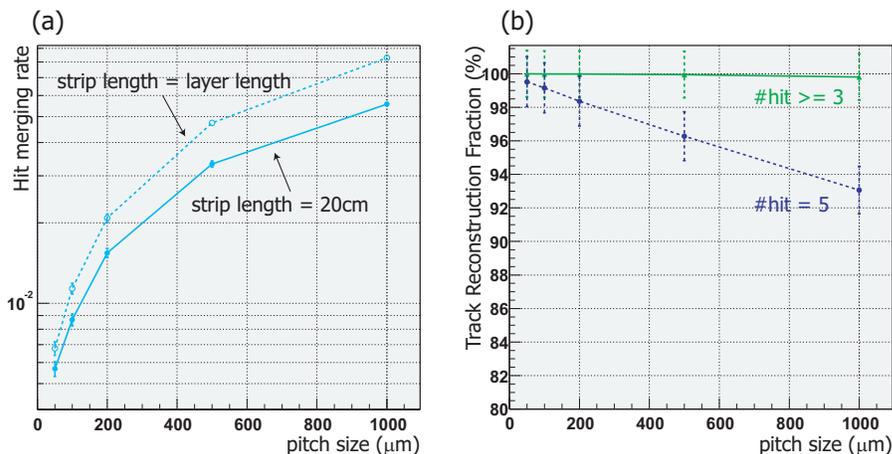}
}
\vskip -0.5cm
\caption{Distributions of 
(a)the number of tracks traversing the same  $r-\phi$ strip 
and (b) the track reconstruction efficiency only using $r-\phi$ strips, 
for various pitch sizes. }
\label{fig:3}
\end{figure}
\begin{figure}
\epsfysize6cm
\center{
\epsfbox{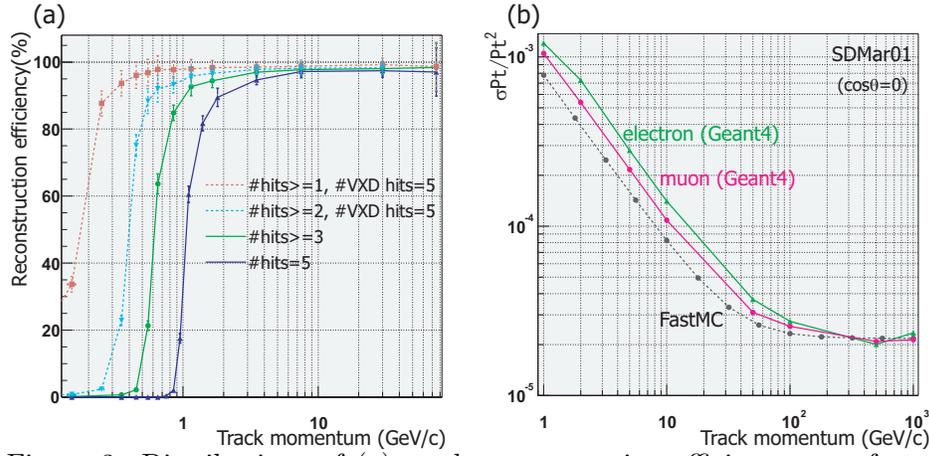}
}
\vskip -0.5cm
\caption{Distributions of (a) track reconstruction 
efficiency as a function of the momentum for the central tracker 
only (solid) and for the central tracker plus vertex detector(dashed), 
and (b) the transverse momentum resolution obtained for the central tracker
plus vertex detector.
}
\label{fig:4}
\end{figure}
\begin{figure}
\epsfysize6.5cm
\center{
\epsfbox{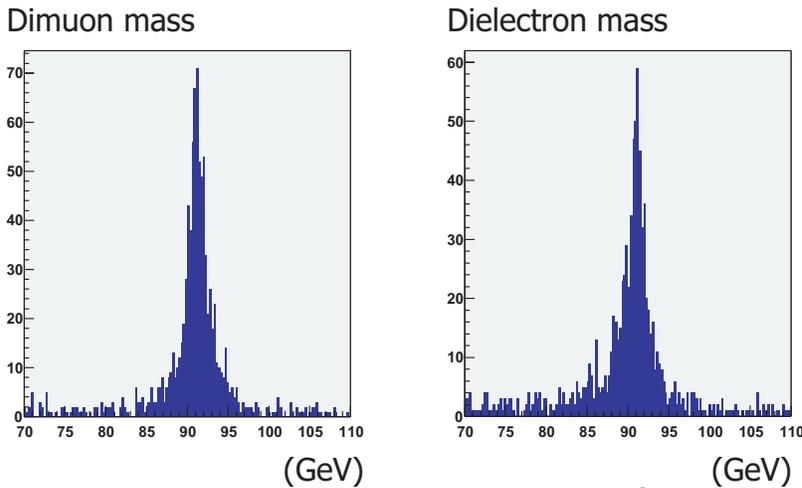}
}
\vskip -0.5cm
\caption{Reconstructed mass distributions for
$Z^0 \rightarrow \mu^+\mu^-$ (left) and 
$Z^0 \rightarrow e^+e^-$ (right) decays,
in $e^+e^-\rightarrow Z^0H^0$ events .
}
\label{fig:5}
\end{figure}
Figure~\ref{fig:3}(a) shows the multiple-hit  rate
in a strip
for various pitch sizes. 
The hits belonging to the same strip is indistinguishable 
if we use only $r-\phi$ information.
For the pitch size  less than 200$\mu$m, only about 1\% of $r-\phi$ strips
have more than one hits.
Figure~\ref{fig:3}(b) shows the track reconstruction efficiency estimated using the $r-\phi$ information {\it only}.
Even using only $r-\phi$ hits we find that 
the reconstruction efficiency is greater  than 99\% assuming we  
we reconstruct tracks with a minimum of 3  layers.
If we require a minimum of 5 layers for track reconstruction, 
we lose the efficiency because some hits are merged for larger pitch sizes.
(This loss, however, can be recovered using the $z$ information provided by  stereo strips.)

Figure~\ref{fig:4}(a) shows the track reconstruction efficiency
as a function of the track momentum, for using the central tracker
only (solid lines) and using the central tracker and vertex detector
(dashed lines). 
Because the solenoid field is high (5 Tesla), 
the presence of the vertex detector is important 
to reconstruct the low momentum tracks.
Figure~\ref{fig:4}(b) shows the transverse 
momentum resolution as a function of the track momentum
obtained using the central tracker and vertex detector.
Here the single track is generated at a zero dip angle (perpendicular  to
the beam axis).
To further study  the  momentum resolution, 
we reconstruct the $Z$ mass, in $e^+e^-\rightarrow Z^0H^0$, 
$Z^0 \rightarrow l^+l^-$ ($l =$ $e$ or $\mu$) events. 
Figure~\ref{fig:5} shows dimuon and dielectron mass distributions obtained using a full simulation.
The $Z$ mass resolution is 1.7 GeV and 1.9 GeV  
for $Z\to \mu^+\mu^-$ and $e^+e^-$, respectively.

We note that
it is important to
use a realistic track-finding algorithm for further study.
In order to estimate the accelerator
background, 
the beam delivery system (BDS) simulation 
is also important.
Using  a newly developed BDS simulation package, 
LCBDS\cite{LCBDS}, based on GEANT4, we plan to study  the effects due to
IR backgrounds.

\end{document}